\documentclass[a4paper,11pt,centertags,reqno,psamsfonts]{amsart}

\usepackage[headings]{fullpage}
\usepackage{setspace}
\usepackage{letltxmacro}
\usepackage{tablefootnote}
\usepackage[inline]{enumitem}
\usepackage{pifont}
\usepackage{array}
\usepackage{fancybox}
\usepackage{hyperref}
\usepackage{ifthen}
\usepackage{graphicx}
\usepackage{subfigure}
\usepackage{amsmath}
\usepackage{amsthm}
\usepackage{amssymb}				
\usepackage{mathtools}
\usepackage{mathrsfs}
\usepackage{multirow}
\usepackage{stmaryrd}
\usepackage{yhmath}
\usepackage{accents}
\usepackage{centernot}
\usepackage{tikz}
\usepackage{tikz-cd}
\usepackage{xfrac}
\usepackage[all]{xy}
\usepackage[round,comma,authoryear,sort&compress]{natbib}

\graphicspath{{Figures/}}

\bibpunct{(}{)}{,}{a}{}{,}

\setlist[itemize]{leftmargin=*,topsep=1ex,itemsep=0ex}
\setlist[enumerate]{leftmargin=*,topsep=1ex,itemsep=0ex}

\allowdisplaybreaks[4]

\makeatletter
\let\oldr@@t\r@@t
\def\r@@t#1#2{%
\setbox0=\hbox{$\oldr@@t#1{#2\,}$}\dimen0=\ht0
\advance\dimen0-0.2\ht0
\setbox2=\hbox{\vrule height\ht0 depth -\dimen0}%
{\box0\lower0.4pt\box2}}
\LetLtxMacro{\oldsqrt}{\sqrt}
\renewcommand*{\sqrt}[2][\ ]{\oldsqrt[#1]{#2}}
\makeatother

\theoremstyle{plain}
\newtheorem{theorem}{Theorem}

\newtheorem{proposition}[theorem]{Proposition}

\theoremstyle{definition}
\newtheorem{definition}[theorem]{Definition}

\theoremstyle{remark}

\numberwithin{theorem}{section}
\numberwithin{equation}{section}
\numberwithin{figure}{section}
\numberwithin{table}{section}

\newcommand{\R}{\mathbb{R}}

\newcommand{\func}[3]{#1:#2\rightarrow #3}

\newcommand{\C}[2]{
\ifthenelse{#1=0 \and #2=0}{\textsf{\upshape C}}
{\ifthenelse{#2=0}{\textsf{\upshape C}^{#1}}
{\textsf{\upshape C}^{#1,#2}}}
}

\newcommand{\e}{\mathrm{e}}
\DeclareMathOperator{\leb}{leb}
\renewcommand{\d}{\mathrm{d}}

\DeclareMathOperator{\BigO}{O}

\newcommand{\SigAlg}[1]{\mathscr{#1}}

\newcommand{\Borel}{\mathscr{B}}

\newcommand{\Filt}[1]{\mathfrak{#1}}

\newcommand{\E}{\textsf{\upshape E}}
\newcommand{\Var}{\textsf{\upshape Var}}

\renewcommand{\P}{\textsf{\upshape P}}
\newcommand{\Prob}[1]{\textsf{\upshape{#1}}}
\newcommand{\ind}[1]{\mathbf{1}_{#1}}

\newcommand{\<}{\langle}
\renewcommand{\>}{\rangle}

\newcommand{\Lint}{\mathsf{L}}

\begin{document}
\title{Arbitrage Problems with Reflected Geometric Brownian Motion}
\thanks{We thank Johannes Ruf for valuable discussions that fine-tuned the arguments in this article. We also thank Claudio Fontana several insightful comments on a previous draft.}

\author{Dean Buckner}
\author{Kevin Dowd}
\author{Hardy Hulley}

\address{Dean Buckner\\
The Eumaeus Project\\
London\\
United Kingdom}
\email{d.e.buckner@eumaeus.org}

\address{Kevin Dowd\\
Durham University Business School\\
Mill Hill Lane\\
Durham DH1 3LB\\
United Kingdom}
\email{kevin.dowd@durham.ac.uk}

\address{Hardy Hulley\\
Finance Discipline Group\\
University of Technology Sydney\\
P.O. Box 123\\
Broadway, NSW 2007\\
Australia}
\email{hardy.hulley@uts.edu.au}



\date{\today}

\begin{abstract}
Contrary to the claims made by several authors, a financial market model in which the price of a risky security follows a reflected geometric Brownian motion is not arbitrage-free. In fact, such models violate even the weakest no-arbitrage condition considered in the literature. Consequently, they do not admit num\'eraire portfolios or equivalent risk-neutral probability measures, which makes them totally unsuitable for contingent claim valuation. Unsurprisingly, the published option pricing formulae for such models violate textbook no-arbitrage bounds.
\end{abstract}

\nocite{Fon15}
\nocite{MW21}
\nocite{FJW20}

\maketitle

\section{Introduction}
In a seminal paper, \citet{Sko61} studied the problem of inserting an instantaneously reflecting boundary into the state space of a one-dimensional It\^o diffusion. The resulting process is described by an SDE that contains a new term---called a reflection term---that controls reflection off the boundary, along with the usual drift and diffusion terms. In addition to proving existence and uniqueness results for the solution to this SDE, \citet{Sko61} derived an explicit expression for the reflection term. Today we recognise the reflection term as the local time of the diffusion at the reflecting boundary.

Several studies have considered financial market models in which the price of a risky security follows a reflected geometric Brownian motion (RGBM). Such a process is obtained by applying \citeauthor{Sko61}'s~\citeyearpar{Sko61} construction to a vanilla geometric Brownian motion, causing it to reflect off a lower boundary. For example, \citet{Vee13}, \citet{Her15}, \citet{NS16} and \citet{HZ17} have used RGBMs to model exchange rates constrained by central bank target zone policies, while \citet{GP00} and \citet{KSW10} modelled the value of an investment fund with a capital guarantee as an RGBM. Recently, \citet{Tho21} modelled house prices as an RGBM, under the assumption that the government will support the property market if prices fall enough. Related models, where the price of a risky security is constrained by reflecting boundaries but the underlying dynamics is not geometric Brownian motion, have been studied by \citet{FJW20} and \citet{MW21}.

\citet{Vee08} claimed that the RGBM model does not offer any arbitrage opportunities. He justified this claim by noting that the security price follows a continuous process and that the time spent by it on the reflecting boundary has Lebesgue measure zero. Based on those observations, he reasoned that arbitrageurs cannot generate riskless profits by purchasing the security when its price reaches the boundary and selling it when its price is reflected off the boundary. On the strength of this heuristic argument, he concluded that the RGBM model is arbitrage-free and must therefore admit an equivalent risk-neutral probability measure. He then obtained an expression for the density of the putative equivalent risk-neutral probability measure and used it to derive pricing formulae for European puts and calls. Following an amendment to the put pricing formula by \citet{HV13}, and with some modifications to cater for dividends, these option pricing formulae have been used by \citet{Vee13}, \citet{Her15}, \citet{HZ17} and \citet{Tho21}.

In this paper, we show that \citeauthor{Vee08}'s~\citeyearpar{Vee08} argument is unsound and his claim that the RGBM model is arbitrage-free is incorrect. In fact, we demonstrate that the model fails to satisfy even the weakest no-arbitrage condition considered in the literature. Consequently, it does not admit a num\'eraire portfolio or an equivalent risk-neutral probability measure. These deficiencies make the RGBM model unsuitable for contingent claim valuation and undermine the validity of the option pricing formulae in the articles cited above. In fact, those formulae are shown to behave quite pathologically when the price of the risky security is close to the reflecting boundary.

We begin, in Section~\ref{Sec2}, with a brief overview of a weak no-arbitrage condition, in the simplified setting of a financial market comprising a bank account and a single non-dividend-paying stock. The primary purpose of this overview is to serve as a roadmap for our subsequent analysis of the RGBM model. In particular, we derive a necessary and sufficient characterisation of the aforementioned no-arbitrage condition that subsequently allows us to pinpoint the exact source of the arbitrage trouble for the RGBM model.

We analyse the RGBM model in Section~\ref{Sec3}. After setting it up and cataloguing its basic properties, we construct a trading strategy that violates the weak no-arbitrage condition studied in Section~\ref{Sec2}, and explain how this result is intimately related to the reflecting behaviour of the stock price in the model. The weak no-arbitrage failure of the RGBM model means that it does not admit an equivalent risk-neutral probability measure, invalidating the risk-neutral approach to contingent claim valuation. Consequently, the published option pricing formulae for the model have no theoretical justification. We examine those formulae in detail, demonstrating that they violate textbook no-arbitrage bounds, and we explain those violations in terms of the reflecting behaviour of the stock price.

In summary, this paper makes two significant contributions. First, it establishes that the RGBM model provides an interesting and non-trivial example of a financial market model that violates a weak no-arbitrage condition, and relates this failure to the characteristics of the model.\footnote{Our analysis could potentially be extended to other financial market models with reflecting boundaries, such as the models studied by \citet{FJW20} and \citet{MW21}.} The second contribution is to highlight the erroneous reasoning about the arbitrage properties of the RGBM model in \citet{Vee08}, which has led to the publication of invalid option pricing formulae by \citet{Vee08,Vee13}, \citet{HV13}, \citet{Her15}, \citet{HZ17} and \citet{Tho21}. Of particular concern is \citet{Tho21}, which recommends the RGBM model to practitioners as a valuation framework for no-negative equity guarantees, a large and important class of insurance products.
\section{A Weak No-Arbitrage Condition and Its Consequences}
\label{Sec2}
This section develops the theoretical prerequisites for our analysis of the reflected geometric Brownian motion model in Section~\ref{Sec3}. We begin by introducing a class of continuous financial market models that is general enough to encompass the reflected geometric Brownian motion model as a particular example. In the context of this modelling framework, we then formulate the so-called \emph{structure condition}, which is expressed in terms of the characteristics of a model, before introducing a weak no-arbitrage condition, known as \emph{no increasing profit}. Our main result establishes that the structure condition and the no-increasing profit condition are equivalent in our setting. Finally, we summarise the economic and modelling consequences of a failure of the no increasing profit condition.

\subsection{A General Modelling Framework}
Throughout this paper, we assume the existence of a filtered probability space $(\Omega,\SigAlg{F},\Filt{F},\P)$, whose filtration $\Filt{F}=(\SigAlg{F}_t)_{t\geq 0}$ satisfies the usual conditions of completeness and right-continuity. It is understood that all random variables and stochastic processes are defined on this space and all stochastic processes are adapted to its filtration. Given a continuous semimartingale $X$, we shall write $\Lint(X)$ to denote the family of predictable processes $\varphi$, such that the stochastic integral $\int_0^\cdot\varphi_s\,\d X_s$ exists.

Consider a financial market comprising a risk-free security and a single risky security. For convenience, we shall refer to the former as a bank account and to the latter as a stock. The value $B$ of the bank account is given by $B_t\coloneqq\e^{rt}$, for all $t\geq 0$, where $r\geq 0$ is a continuously compounding risk-free interest rate. The stock price $S$ is determined by the SDE
\begin{equation}
\label{eqSec2:StockPrice}
\d S_t=S_t\,\d A_t+S_t\,\d M_t
\end{equation} 
for all $t\geq 0$, with initial value $S_0>0$, where $A$ is a continuous finite variation process and $M$ is a continuous local martingale. For convenience, we shall assume that $\<M\>_t<\infty$, for all $t\geq 0$, which ensures that the stock price is strictly positive, by virtue of the law of large numbers for local martingales \citep[see][Exercise~V.1.16]{RY99}.

Investors in the market described above are able to trade self-financing portfolios comprising the bank account and the stock. The following definition makes this concept precise.

\begin{definition}
\label{defSec2:TradStrat}
A \emph{trading strategy} is a predictable process $\xi\in\Lint(S)$. Given a trading strategy $\xi\in\Lint(S)$ and an initial endowment $v\geq 0$, the stochastic process $V^{v,\xi}$, determined by the SDE
\begin{equation}
\label{eqdefSec2:TradStrat}
\d V^{v,\xi}_t=\frac{V^{v,\xi}_t-\xi_tS_t}{B_t}\,\d B_t+\xi_t\,\d S_t
=r(V^{v,\xi}_t-\xi_tS_t)\,\d t+\xi_tS_t\,\d A_t+\xi_tS_t\,\d M_t,
\end{equation}
for all $t\geq 0$, with initial value $V^{v,\xi}_0=v$, is called the \emph{portfolio value} generated by $v$ and $\xi$.
\end{definition}

Given a trading strategy $\xi\in\Lint(S)$ and an initial endowment $v\geq 0$, we interpret $V^{v,\xi}_t$ as the value at time $t\geq 0$ of a self-financing portfolio that holds $\xi_s$ shares of the stock at each time $s\in[0,t]$, with initial value $V^{v,\xi}_0=v$. The SDE \eqref{eqdefSec2:TradStrat} follows from the self-financing requirement that the portfolio must hold $\sfrac{(V^{v,\xi}_t-\xi_tS_t)}{B_t}$ units of the bank account at time $t\geq 0$, if it holds $\xi_t$ shares of the stock at that time.

\subsection{The Structure Condition}
Following \citet[Proposition~II.2.9]{JS03}, there exists a continuous (and hence predictable) increasing process $G$, such that
\begin{equation}
\label{eqSec2:PredChar}
rt=\int_0^t\rho_s\,\d G_s,\qquad 
A_t=\int_0^tb_s\,\d G_s\qquad\text{and}\qquad
\<M\>_t=\int_0^ta_s^2\,\d G_s,
\end{equation}
for all $t\geq 0$, where $\rho$, $b$ and $a$ are some predictable processes. One possible choice for $G$ is obtained by setting $G_t\coloneqq t+\Var(A)_t+\<M\>_t$, for all $t\geq 0$. The next definition formulates the so-called \emph{structure condition} in terms of the processes $\rho$, $b$ and $a$.

\begin{definition}
\label{defSec2:SC}
The financial market satisfies the \emph{structure condition} if there is a predictable process $\vartheta$, such that
\begin{equation}
\label{eqdefSec2:SC}
b_t(\omega)-\rho_t(\omega)=\vartheta_t(\omega)a_t^2(\omega),
\end{equation}
for $\P\otimes G\text{-a.a}\;(\omega,t)\in\Omega\times\R_+$.
\end{definition}

The structure condition was first identified by \citet{Sch95}, who formulated it somewhat differently. To demonstrate that the two formulations are equivalent, let $\hat{S}$ denote the discounted stock price, defined by $\hat{S}_t\coloneqq\sfrac{S_t}{B_t}=\e^{-rt}S_t$, for all $t\geq 0$. It satisfies the SDE
\begin{equation*}
\d\hat{S}_t=-r\hat{S}_t\,\d t+\hat{S}_t\,\d A_t+\hat{S}_t\,\d M_t,
\end{equation*}
for all $t\geq 0$, with $\hat{S}_0=S_0>0$, by an application of It\^o's formula. Next, define the continuous finite variation process $\hat{A}$ and the continuous local martingale $\hat{M}$, by setting
\begin{equation}
\label{eqSec2:A^&B^}
\hat{A}_t\coloneqq-\int_0^tr\hat{S}_u\,\d u+\int_0^t\hat{S}_u\,\d A_u
\qquad\text{and}\qquad
\hat{M}_t\coloneqq\int_0^t\hat{S}_u\,\d M_u,
\end{equation}
for all $t\geq 0$. If the structure condition holds, for some predictable process $\vartheta$ satisfying \eqref{eqdefSec2:SC}, then
\begin{equation*}
\begin{split}
\d\hat{A}_t(\omega)=-r\hat{S}_t(\omega)\,\d t+\hat{S}_t(\omega)\,\d A_t(\omega)
&=\bigl(b_t(\omega)-\rho_t(\omega)\bigr)\hat{S}_t(\omega)\,\d G_t(\omega)\\
&=\vartheta_t(\omega)a_t^2(\omega)\hat{S}_t(\omega)\,\d G_t(\omega)\\
&=\vartheta_t(\omega)\hat{S}_t(\omega)\,\d\<M\>_t(\omega)\\
&=\frac{\vartheta_t(\omega)}{\hat{S}_t(\omega)}\,\d\<\hat{M}\>_t(\omega),
\end{split}
\end{equation*}
for $\P\otimes G\text{-a.a}\;(\omega,t)\in\Omega\times\R_+$. On the other hand, if there is a predictable process $\lambda$, such that
\begin{equation}
\label{eqSec2:AltSC}
\d\hat{A}_t(\omega)=\lambda_t(\omega)\,\d\<\hat{M}\>_t(\omega),
\end{equation}
for $\P\otimes G\text{-a.a}\;(\omega,t)\in\Omega\times\R_+$, then the same reasoning shows that the structure condition holds, with the predictable process $\vartheta\coloneqq\lambda\hat{S}$ satisfying \eqref{eqdefSec2:SC}. So, the structure condition is equivalent to the existence of a predictable process $\lambda$ that satisfies \eqref{eqSec2:AltSC}, which is how \citet{Sch95} originally formulated it.

The reformulation of the structure condition above offers a useful insight that will help us interpret the arbitrage properties of the reflected geometric Brownian motion model, studied in Section~\ref{Sec3}. First, note that the finite variation processes $\hat{A}$ and $\<\hat{M}\>$ may be regarded as (possibly signed) random measures on $(\R_+,\Borel(\R_+))$. With that interpretation in mind, \eqref{eqSec2:AltSC} states that $\hat{A}\ll\<\hat{M}\>$, with predictable density $\lambda$. This means that if $\int_0^\infty\ind{U}\,\d\<\hat{M}\>_t(\omega)=0$ then $\int_0^\infty\ind{U}\,\d\hat{A}_t(\omega)=0$, for $\P\text{-a.a.}\;\omega\in\Omega$ and any Borel-measurable set $U\in\Borel(\R_+)$. In particular, if there are measurable subsets of $\R_+$ over which the sample paths of $\hat{A}$ increase or decrease but the sample paths of $\<\hat{M}\>$ remain constant, then the structure condition cannot hold.
 
\subsection{The No Increasing Profit Condition}
An \emph{increasing profit} is the strongest form of arbitrage for continuous-time financial market models. This concept of arbitrage was introduced by \citet{KK07a} and thoroughly investigated by \citet{Fon15}. The following definition provides a formulation that is appropriate for our setting.

\begin{definition}
\label{defSec2:NIP}
The financial market admits an \emph{increasing profit} if there is a trading strategy $\xi\in\Lint(S)$, such that
\begin{enumerate}
\item
$V^{0,\xi}$ is non-decreasing; and
\item
$\P(V^{0,\xi}_{\infty-}>0)>0$.
\end{enumerate}
The market satisfies the \emph{no increasing profit (NIP) condition} if no such strategy exists.
\end{definition}

Putting the previous definition into words, a model admits an increasing profit if there is a trading strategy, for which the value of the associated portfolio with an initial endowment of zero is a non-decreasing process with a positive probability of ultimately becoming strictly positive. Such arbitrage opportunities are unequivocally pathological, since they offer investors a chance of making something from nothing, without incurring any risk. Equilibrium asset prices cannot exist in a model that offers such opportunities, since there would be insatiable demand for the portfolios that exploit them. Consequently, a viable financial market model must satisfy the NIP condition.

The next theorem reveals that the NIP condition is equivalent to the structure condition. \citet[Theorem~3.1]{Fon15} proved essentially the same result, but the formulation and proof in our setting are quite different and the proof is sufficiently instructive to merit inclusion. \citet[Lemma~1.4.6]{KS91} and \citet[Theorem~3.5]{DS95b} proved related results in the ``only iff'' direction. 

\begin{theorem}
\label{thmSec2:NIP<=>SC}
The NIP condition is satisfied if and only if the structure condition is satisfied.
\end{theorem}
\begin{proof}
($\Rightarrow$)~Suppose the NIP condition holds. Consider a portfolio with initial endowment $v=1$ that implements a trading strategy $\xi\in\Lint(S)$, specified by
\begin{equation*}
\xi_t\coloneqq\ind{\{a_t=0\}}\bigl(\ind{\{b_t>\rho_t\}}-\ind{\{b_t<\rho_t\}}\bigr)\frac{V^{1,\xi}_t}{S_t},
\end{equation*}
for all $t\geq 0$. In that case, the quadratic variation of the local martingale $\int_0^\cdot\xi_uS_u\,\d M_u$ satisfies
\begin{equation*}
\begin{split}
\biggl\<\int_0^\cdot\xi_uS_u\,\d M_u\biggr\>_t
=\int_0^t\xi_u^2S_u^2\,\d\<M\>_u
&=\int_0^t\xi_u^2a_u^2S_u^2\,\d G_u\\
&=\int_0^t\ind{\{a_u=0\}}\bigl(\ind{\{b_u>\rho_u\}}-\ind{\{b_u<\rho_u\}}\bigr)^2a_u^2(V^{1,\xi}_u)^2\,d G_u
=0,
\end{split}
\end{equation*}
for all $t\geq 0$. This implies that $\int_0^\cdot\xi_uS_u\,\d M_u=0$, since a local martingale whose quadratic variation is zero must remain constant. Using differential notation, the latter condition can be expressed as $\xi_tS_t\,\d M_t=0$, for all $t\geq 0$. Hence, \eqref{eqdefSec2:TradStrat} gives
\begin{equation*}
\begin{split}
\d V^{1,\xi}_t&=r(V^{1,\xi}_t-\xi_tS_t)\,\d t+\xi_tS_t\,\d A_t\\
&=rV^{1,\xi}_t\,\d t+\xi_t(b_t-\rho_t)S_t\,\d G_t\\
&=rV^{1,\xi}_t\,\d t+\ind{\{a_t=0\}}\bigl(\ind{\{b_t>\rho_t\}}-\ind{\{b_t<\rho_t\}}\bigr)(b_t-\rho_t)V^{1,\xi}_t\,\d G_t\\
&=rV^{1,\xi}_t\,\d t+\ind{\{a_t=0\:\text{and}\:b_t\neq\rho_t\}}|b_t-\rho_t|V^{1,\xi}_t\,\d G_t,
\end{split}
\end{equation*}
for all $t\geq 0$. The solution to this equation is
\begin{equation*}
V^{1,\xi}_t=\exp\biggl(rt+\int_0^t\ind{\{a_s=0\;\text{and}\;b_s\neq\rho_s\}}|b_s-\rho_s|\,\d G_s\biggr),
\end{equation*}
for all $t\geq 0$. Now consider another trading strategy $\tilde{\xi}\in\Lint(S)$, with initial endowment $\tilde{v}=0$, comprising a long position in $\xi$ and a short position in the bank account. The value of the associated portfolio is given by
\begin{equation*}
V^{0,\tilde{\xi}}_t=V^{1,\xi}_t-B_t
=e^{rt}\biggl(\exp\biggl(\int_0^t\ind{\{a_s=0\;\text{and}\;b_s\neq\rho_s\}}|b_s-\rho_s|\,\d G_s\biggr)-1\biggr),
\end{equation*}
for all $t\geq 0$. Since the integrand in this expression is non-negative, it follows that $V^{0,\tilde{\xi}}$ is a non-decreasing process. This implies that $V^{0,\tilde{\xi}}_{\infty-}=0$, by virtue of the assumption that the NIP condition holds, which is to say that
\begin{equation*}
\int_0^\infty\ind{\{a_t=0\;\text{and}\;b_t\neq\rho_t\}}|b_t-\rho_t|\,\d G_t=0.
\end{equation*}
From this we conclude that $b_t(\omega)=\rho_t(\omega)$, for $\P\otimes G\text{-a.a}\;(\omega,t)\in\{a_\cdot(\,\cdot\,)=0\}$, whence the predictable process $\vartheta$, defined by
\begin{equation*}
\vartheta_t\coloneqq\ind{\{a_t\neq 0\}}\frac{b_t-\rho_t}{a_t^2},
\end{equation*}
for all $t\geq 0$, satisfies \eqref{eqdefSec2:SC} and verifies the structure condition.
\vspace{2mm}\noindent\newline
($\Leftarrow$)~Suppose the structure condition is satisfied and let $\xi\in\Lint(S)$ be a trading strategy, for which the value $V^{0,\xi}$ of the associated portfolio with zero initial endowment is non-decreasing. It follows that $V^{0,\xi}$ is a finite variation process, which implies that
\begin{equation*}
\<V^{0,\xi}\>_t=\biggl\<\int_0^\cdot\xi_uS_u\,\d M_u\biggr\>_t
=\int_0^t\xi_u^2S_u^2\,\d\<M\>_u
=\int_0^t\xi_u^2S_u^2a_u^2\,\d G_u
=0,
\end{equation*}
for all $t\geq 0$, whence $a_t(\omega)=0$, for $\P\otimes G\;(\omega,t)\in\{\xi_\cdot(\,\cdot\,)\neq 0\}$. Since the structure condition holds, by assumption, it follows from \eqref{eqdefSec2:SC} that $b_t(\omega)=\rho_t(\omega)$, for $\P\otimes G\text{-a.a}\;(\omega,t)\in\{\xi_\cdot(\,\cdot\,)\neq 0\}$. Next, observe that $\<\int_0^\cdot\xi_uS_u\,\d M_u\>=0$ also implies that $\int_0^\cdot\xi_uS_u\,\d M_u=0$, since a local martingale with zero quadratic variation remains constant. This can be expressed as $\xi_tS_t\,\d M_t=0$, for all $t\geq 0$. Consequently, \eqref{eqdefSec2:TradStrat} gives
\begin{equation*}
\d V^{0,\xi}_t=r(V^{0,\xi}_t-\xi_tS_t)\,\d t+\xi_tS_t\,\d A_t
=rV^{0,\xi}_t\,\d t+\xi_tS_t(b_t-\rho_t)\,\d G_t,
\end{equation*}
for all $t\geq 0$. Since we have established that $b_t(\omega)=\rho_t(\omega)$, for $\P\otimes G\text{-a.a}\;(\omega,t)\in\{\xi_\cdot(\,\cdot\,)\neq 0\}$, it follows from the expression above that $\d V^{0,\xi}_t=rV^{0,\xi}_t\,\d t$, for all $t\geq 0$, whence $V^{0,\xi}_t=V^{0,\xi}_0\e^{rt}=0$, since $V^{0,\xi}_0=0$. Consequently, $V^{0,\xi}_{\infty-}=0$,  implying that $\P(V^{0,\xi}_{\infty-}>0)=0$. In summary, we have demonstrated that any trading strategy that satisfies the first condition in Definition~\ref{defSec2:NIP} cannot satisfy the second condition. Hence, the NIP condition holds.
\end{proof}

Inspection of \eqref{eqdefSec2:SC} reveals that the structure condition fails if and only if there are periods during which the process $a$ is zero and processes $b$ and $\rho$ assume different values. Whenever that happens, the market effectively contains two risk-free securities with different rates of return. During such periods, the trading strategy constructed in the first part of the proof of Thoerem~\ref{thmSec2:NIP<=>SC} generates an increasing profit by borrowing at the lower rate and investing the proceeds at the higher rate. In light of Theorem~\ref{thmSec2:NIP<=>SC}, the NIP condition may be interpreted economically as stating that the market never contains two risk-free securities with different rates of return.

\subsection{Contingent Claim Valuation}
In order for a model to be suitable for contingent claim valuation, it should admit either a \emph{num\'eraire portfolio} or an \emph{equivalent risk-neutral probability measure}. Informally, a num\'eraire portfolio in our modelling framework is a well-behaved trading strategy $\xi^*$, whose portfolio value $V^{1,\xi^*}$, with initial endowment $v=1$, provides a natural benchmark against which the value of every other well-behaved portfolio can be measured. More formally, the \emph{benchmarked processes} $\sfrac{B}{V^{1,\xi^*}}$ and $\sfrac{S}{V^{1,\xi^*}}$ are required to be supermartingales. On the other hand, an equivalent risk-neutral probability measure in our setting is an equivalent probability measure $\Prob{Q}\sim\P$, such that the discounted stock price $\hat{S}\coloneqq\sfrac{S}{B}$ is a local martingale under $\Prob{Q}$. If a model admits a num\'eraire portfolio, then the \emph{benchmark approach} to contingent claim valuation \citep[see][]{PH06} can be used, while traditional \emph{risk-neutral valuation} can be used if it admits an equivalent risk-neutral probability measure.

Both the existence of well-behaved num\'eraire portfolio and the existence of an equivalent risk-neutral probability measure can be characterised in terms of no-arbitrage conditions. In more general settings than ours, \citet{KK07a} established that the existence of a well-behaved num\'eraire portfolio is equivalent to the \emph{no unbounded profit with bounded risk (NUPBR)} condition, while \citet{DS94b} famously demonstrated that the \emph{no free lunch with vanishing risk (NFLVR)} condition is necessary and sufficient for the existence of an equivalent risk-neutral probability measure. Based on those results, a model that does not satisfy NUPBR or NFLVR cannot be used for pricing options and other contingent claims.

Crucially, both NUPBR and NFLVR are stronger conditions that NIP.\footnote{The logical dependencies between the various no-arbitrage conditions for continuous financial market models are explained in the surveys by \citet[Chapter~1]{Hul10b} and \citet{Fon15}.} Consequently, a model that does not satisfy the NIP condition will fail to satisfy the NUPBR and NFLVR conditions as well, which means that it will not admit a well-behaved num\'eraire portfolio or an equivalent risk-neutral probability measure. Such a model is unsuitable for contingent claim valuation. This is unsurprising, since we have already argued that a model cannot sustain equilibrium asset prices if it does not satisfy the NIP condition.

\section{A Reflected Geometric Brownian Motion Model}
\label{Sec3}
\citet{Vee08} proposed a modification of the \citet{BS73} model, in which the price of a risky asset is a geometric Brownian with a reflecting lower boundary. A related model was employed by \citet{GP00}, and \citet{KSW10} for the value of an investment fund with a capital guarantee, while \citet{Vee13}, \citet{NS16}, \citet{Her15}, and \citet{HZ17} used similar models for exchange rates constrained by target zones. Finally, \citet{Tho21} recently used reflected geometric Brownian motion to model house prices protected by a government guarantee. This section provides a careful analysis of the arbitrage properties of the reflected geometric Brownian motion model. We begin with a rigorous formulation of the model, before describing its basic properties. Next, we demonstrate that it violates both the no increasing profit condition and the structure condition, before analysing those failures in terms of the reflecting behaviour of the asset price in the model. Finally, we demonstrate that the option pricing formulae for the model, presented in several of the previously cited studies, exhibit numerous pathologies, which are once again related to the reflecting behaviour of the asset price in the model.
 
\subsection{Reflected Geometric Brownian Motion}
\citet{Sko61} showed that the SDE
\begin{equation}
\label{eqSec3:RGBM}
\d S_t=\mu S_t\,\d t+\sigma S_t\,\d W_t+\d L_t,
\end{equation}
for all $t\geq 0$, with a lower reflecting boundary $b>0$ and initial value $S_0>b$, admits a unique strong solution. This solution comprises a pair of processes $S$ and $L$, such that
\begin{subequations}
\label{eqSec3:RGBMSol}
\begin{align}
&S_t\geq b;\label{eqSec3:RGBMSol-a}\\
&\text{$L$ is non-decreasing with $L_0=0$};\label{eqSec3:RGBMSol-b}\\
&\int_0^t\ind{\{S_u>b\}}\d L_u=0;\qquad\text{and}\label{eqSec3:RGBMSol-c}\\
&S_t=S_0+\int_0^t\mu S_u\,\d u+\int_0^t\sigma S_u\,\d W_u+L_t,\label{eqSec3:RGBMSol-d}
\end{align}
\end{subequations}
for all $t\geq 0$, and where the integrals above are well-defined. The process $S$ is called a \emph{reflected geometric Brownian motion (RGBM)} and the process $L$ is called the \emph{reflection term}.

Condition~\eqref{eqSec3:RGBMSol-c} ensures that the value of $L$ does not change while the value of $S$ exceeds $b$, in which case condition~\eqref{eqSec3:RGBMSol-d} ensures that $S$ behaves like a vanilla geometric Brownian motion (GBM). However, as soon as $S$ reaches the boundary $b$, the value of $L$ increases, instantaneously reflecting $S$ back into the interval $(b,\infty)$, after which it behaves like a vanilla GBM once again. It follows that the points of increase of $L$ are limited to times when $S$ visits $b$. It is natural to interpret $L_t$ as the cumulative amount of reflection of $S$ from the boundary $b$, up time $t\geq0$.

In addition to establishing the existence of a unique solution to \eqref{eqSec3:RGBM}, \citet{Sko61} showed that the reflection term is given by
\begin{equation*}
L_t=\sqrt{\frac{\pi}{8}}\int_0^t\ind{\{S_u=b\}}\sigma\,\sqrt{\d u},
\end{equation*}
for all $t\geq 0$, where the integral in this expression can be defined rigorously as a limit of integral sums. There is, however, a more useful interpretation of the reflection term. Define the process $\ell^b$, by setting
\begin{equation}
\label{eqSec3:RGBMLocTime}
\ell^b_t\coloneqq\lim_{\varepsilon\downarrow 0}\frac{1}{\varepsilon}\int_0^t\ind{(b,b+\varepsilon]}(S_u)\,\d\<S\>_u
=\lim_{\varepsilon\downarrow 0}\frac{1}{\varepsilon}\int_0^t\ind{\{b<S_u\leq b+\varepsilon\}}\sigma^2S_u^2\,\d u,
\end{equation}
for all $t\geq 0$. This process, which is known as the \emph{local time} of $S$ at $b$, provides a non-trivial measure of the amount of time $S$ spends in the vicinity of $b$. \citet[Theorem~1.3.1]{Pil14} showed that $L_t=\sfrac{\ell^b_t}{2}$, for all $t\geq 0$, which is to say that the reflection term is just the scaled local time at the reflecting boundary.

\subsection{Model Specification}
We shall now revisit the financial market described in Section~\ref{Sec2}, comprising a risk-free bank account and a non-dividend-paying stock. As before, the value $B$ of the bank account is given by $B_t\coloneqq\e^{rt}$, for all $t\geq 0$, where $r\geq 0$ is the risk-free interest rate. However, following \citet{Vee08}, we now assume that the stock price $S$ follows the RGBM \eqref{eqSec3:RGBM}.

We begin by noting that the RGBM model falls within the scope of the framework introduced in Section~\ref{Sec2}, which ensures that the analysis in that section is fully applicable to it. Indeed, \eqref{eqSec3:RGBMSol-c} allows us to rewrite \eqref{eqSec3:RGBM} as
\begin{equation*}
\d S_t=\mu S_t\,\d t+\sigma S_t\,\d W_t+\frac{S_t}{b}\,\d L_t,
\end{equation*}
for all $t\geq 0$. This equation may in turn be expressed in the form \eqref{eqSec2:StockPrice}, with the continuous finite variation process $A$ and the continuous local martingale $M$ given by
\begin{equation}
\label{eqSec3:A&M}
A_t\coloneqq\mu t+\frac{L_t}{b}
\qquad\text{and}\qquad
M_t\coloneqq\int_0^t\sigma\,\d W_s,
\end{equation}
for all $t\geq 0$. As noted in Section~\ref{Sec2}, there exists a continuous increasing process $G$ satisfying \eqref{eqSec2:PredChar}, for some predictable processes $\rho$, $b$ and $a$. That is to say,
\begin{equation}
\label{eqSec3:PredChar}
r\,\d t=\rho_t\,\d G_t,
\qquad
\mu\,\d t+\frac{\d L_t}{b}=b_t\,\d G_t
\qquad\text{and}\qquad
\sigma^2\,\d t=a_t^2\,\d G_t,
\end{equation}
for all $t\geq 0$. Setting $G_t\coloneqq t+\ell^b_t$, for all $t\geq 0$, provides a convenient choice for the process $G$.

\subsection{Failure of the No Increasing Profit Condition}
The behaviour of the stock price at the reflecting boundary raises concerns about the arbitrage properties of the RGBM model. Indeed, it seems plausible that an arbitrageur could exploit this behaviour by purchasing the stock when its price reaches the boundary and unwinding the position immediately afterwards, as the stock price is reflected off the boundary. The guiding intuition is that, upon reaching the boundary, the behaviour of the stock price is completely predictable over the next instant of time, and therefore arbitrageable. The following proposition confirms this intuition.\footnote{The trading strategy constructed in the proof of Proposition~\ref{propSec3:NIP} is similar in spirit to the trading strategy in \citet[Example~7.1]{Fon15}.}

\begin{proposition}
\label{propSec3:NIP}
The RGBM model violates the NIP condition.
\end{proposition}
\begin{proof}
Define the process $\xi$, by setting $\xi_t\coloneqq\ind{\{S_t=b\}}$, for all $t\geq 0$. Then $\xi$ is predictable, since it is obtained by applying the measurable function $\func{\ind{\{\,\cdot\,=b\}}}{\R}{\R}$ to the continuous (and hence predictable) process $S$. Moreover, since every bounded predictable process is integrable with respect to any semimartingale, it follows that $\xi\in\Lint(S)$, which establishes that $\xi$ is a valid trading strategy. Now, since $\leb\{t\in\R_+\,|\,S_t(\omega)=b\}=0$, for $\P\text{-a.a.}\;\omega\in\Omega$, the properties of the Lebesgue and It\^o integrals and the local time process ensure that
\begin{align*}
\frac{\xi_tS_t}{B_t}\,\d B_t&=\ind{\{S_t=b\}}rS_t\,\d t=0\\
\intertext{and}
\xi_t\,\d S_t&=\ind{\{S_t=b\}}\mu S_t\,\d t+\ind{\{S_t=b\}}\sigma S_t\,\d W_t+\ind{\{S_t=b\}}\,\d L_t=\d L_t,
\end{align*}
for all $t\geq 0$. By substituting these identities into \eqref{eqdefSec2:TradStrat}, we see that the value $V^{0,\xi}$ of a portfolio that implements the strategy $\xi$, with zero initial endowment, satisfies the SDE
\begin{equation}
\label{eqpropSec3:NIP}
\d V^{0,\xi}_t=\frac{V^{0,\xi}_t-\xi_t S_t}{B_t}\,\d B_t+\xi_t\,\d S_t
=rV^{0,\xi}_t\,\d t+\d L_t,
\end{equation}
for all $t\geq 0$. First, since $L$ is non-decreasing, it follows that $V^{0,\xi}$ is non-decreasing as well. Second, since $\d L_t(\omega)>0$, for $\P\otimes G\text{-a.a.}\:(\omega,t)\in\{S_\cdot(\,\cdot\,)=b\}$, it follows that $\ind{\{\tau_b<\infty\}}V^{0,\xi}_t>0$, for all $t\geq\tau_b$, where $\tau_b\coloneqq\inf\{t\geq 0\,|\,S_t=b\}$ denotes the first-passage time of $S$ to $b$. Since the properties of geometric Brownian motion ensure that $\P(\tau_b<\infty)>0$, this implies that $\P(V^{0,\xi}_{\infty-}>0)>0$. Together, these observations confirm that $\xi$ generates an increasing profit.
\end{proof}

\begin{figure}
\subfigure[Sample path of the stock price.]{
\centering
\includegraphics[scale=0.75]{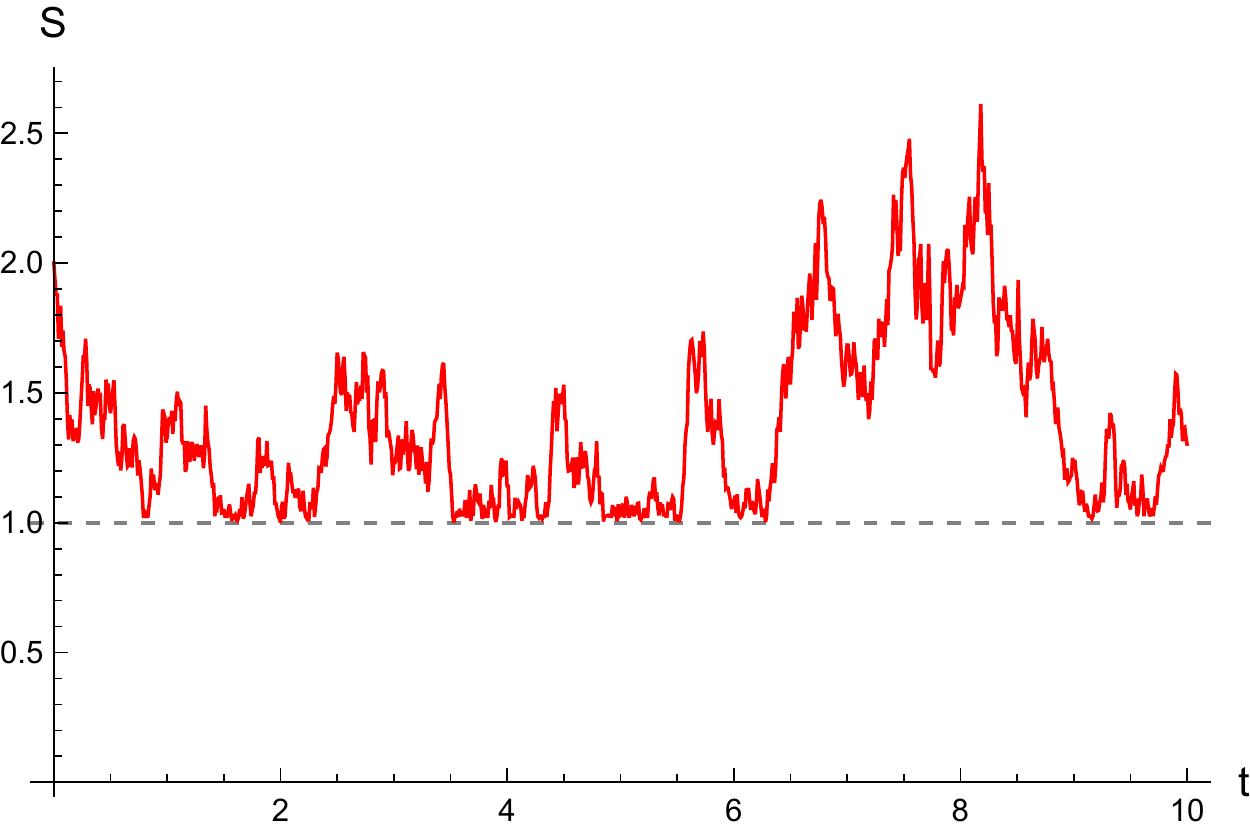}
\label{figSec3:RGBM-S}}
\subfigure[Sample path of the reflection term.]{
\includegraphics[scale=0.75]{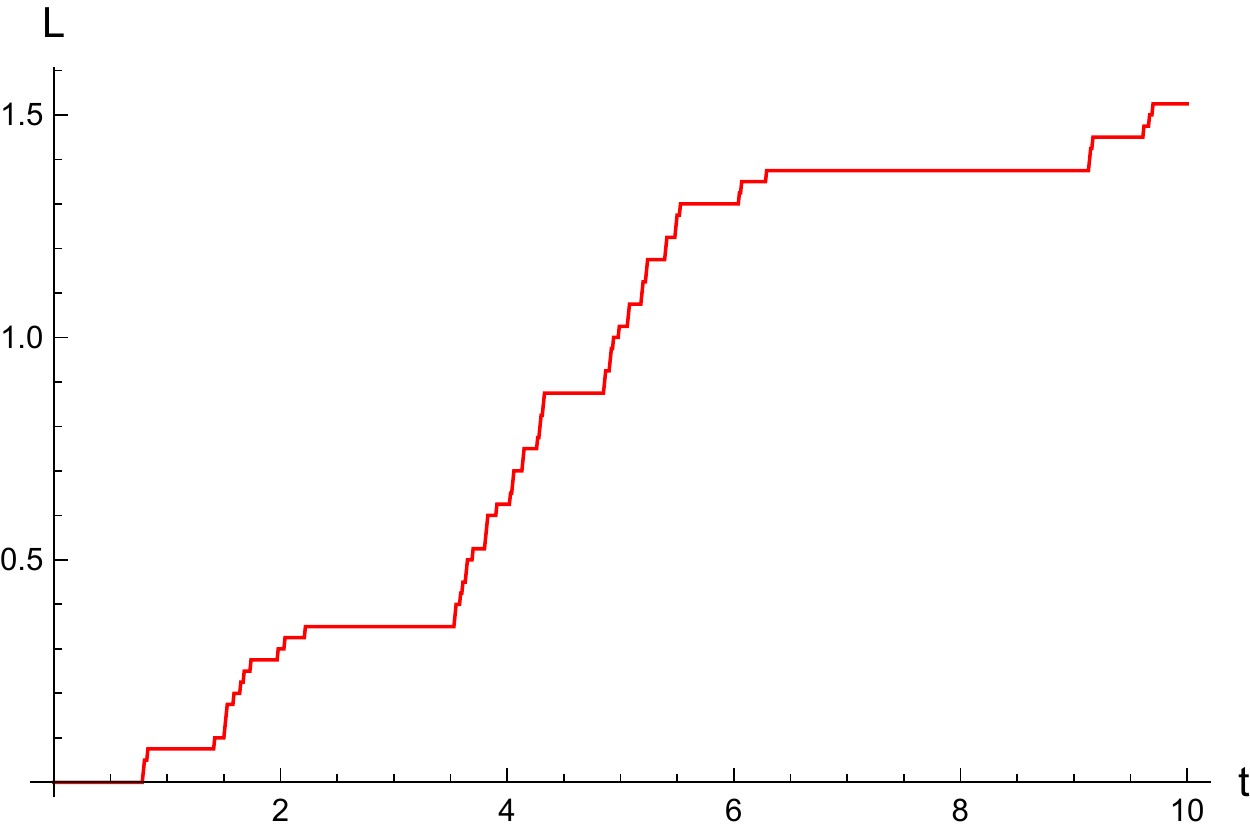}
\label{figSec3:RGBM-L}}
\subfigure[Sample path of the arbitrage strategy.]{
\includegraphics[scale=0.75]{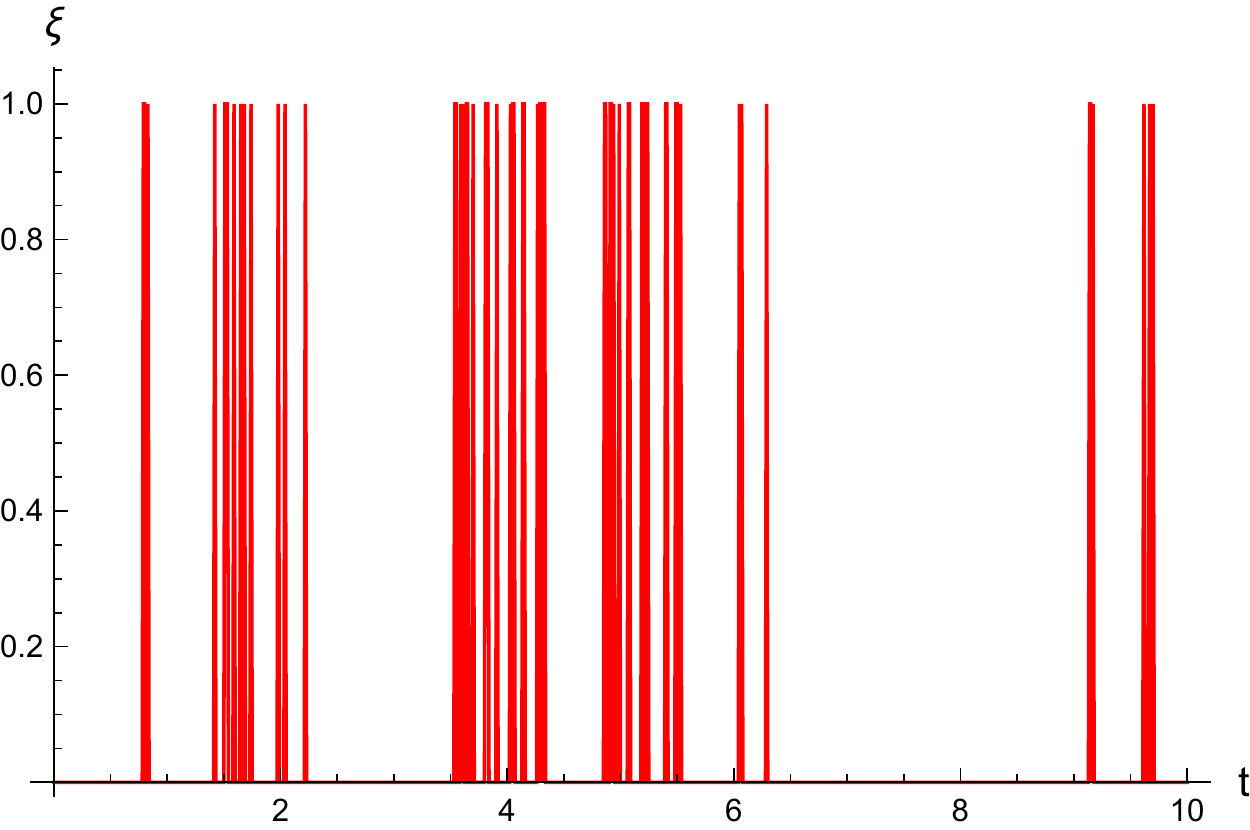}
\label{figSec3:RGBM-xi}}
\caption{Sample path for the stock price and the reflection term determined by \eqref{eqSec3:RGBM}, with $\mu=0$, $\sigma=0.5$, $b=1$ and $S_0=2$, as well as the corresponding sample path for the immediate arbitrage strategy in the proof of Proposition~\ref{propSec3:NIP}.}
\label{figSec3:RGBM}
\end{figure}

Figure~\ref{figSec3:RGBM} plots a simulated path for the stock price and the corresponding path for the reflection term, together with the path for the stock position in the arbitrage portfolio described in the proof of Proposition~\ref{propSec3:NIP}. Figure~\ref{figSec3:RGBM}\subref{figSec3:RGBM-S} illustrates the reflection of the stock price off the boundary, while Figure~\ref{figSec3:RGBM}\subref{figSec3:RGBM-L} illustrates the behaviour of the the reflection term (which is effectively the local time of the stock price at the boundary). Each time the stock price visits the boundary, we see an instantaneous increase in the refection term, which nudges the stock price away from the boundary. Figure~\ref{figSec3:RGBM}\subref{figSec3:RGBM-xi} shows that the stock holding in the arbitrage portfolio switches from zero to one in that instant, before immediately switching back to zero as the stock price is instantaneously reflected off the boundary. In effect, the stock is purchased and immediately sold when its price reaches the boundary, realising a profit equal to the increment in the reflection term in that instant. This profit is deposited in the bank account, so that the value of the arbitrage portfolio at any time is the cumulative sum of the instantaneous profits realised up to that time, plus interest.
 
In the case when the risk-free interest rate is zero, it follows from \eqref{eqpropSec3:NIP} that the value of the arbitrage portfolio is simply the value of the reflection term (or the scaled local time process). In that case, Figure~\ref{figSec3:RGBM}\subref{figSec3:RGBM-L} also illustrates the behaviour of the value of the arbitrage portfolio. We see that the portfolio requires no initial investment of capital, since its initial value is zero. We also see that its value is non-decreasing, which reflects the riskless nature of the strategy as well as the fact that it requires no trading capital to fund margin calls. Moreover, the portfolio value increases instantaneously from zero at the first-passage time of the stock price to the reflecting boundary, and increases again each subsequent time the stock price visits the boundary. Effectively, the arbitrage strategy harvests the local time of the stock price at the boundary as a riskless profit.

\citet{Vee08} claimed that the RGBM model does not admit arbitrage opportunities. He justified this assertion by citing two properties of the stock price in the model. First, since reflection off the boundary is instantaneous, the Lebesgue measure of the time spent there by the stock price is zero. Second, the stock price follows a continuous process. Based on those two observations (both of which are correct), he argued that the model is arbitrage-free and therefore admits an equivalent risk-neutral probability measure. Although Proposition~\ref{propSec3:NIP} refutes this argument, we shall nevertheless dwell on it, because it has been repeated by several other authors.

The main problem is that \citeauthor{Vee08}'s~\citeyearpar{Vee08} argument focuses on the wrong measure of the time spent by the stock price at the reflecting boundary. Although the Lebesgue measure of the time spent there is indeed zero, that is irrelevant. Instead, we should focus on the local time of the stock price at the boundary. Being a continuous process whose paths are of unbounded total variation on compact intervals means that the sample paths of the stock price are extremely irregular. As a result of this irregularity, the stock price spends a non-zero amount of time infinitesimally close to the boundary, each time it visits, even though it spends no time actually on the boundary. The time spent arbitrarily close to the boundary is what the local time process measures. Each time the stock price reaches the boundary, the clock measuring its local time there ticks over and the local time increment is added to the stock price, nudging it away from the boundary. The arbitrage strategy constructed in the proof of Proposition~\ref{propSec3:NIP} exploits this behaviour, by purchasing and immediately selling the stock each time it reaches the boundary, risklessly harvesting the local time increments in the process.

Another problem with \citeauthor{Vee08}'s~\citeyearpar{Vee08} argument is its apparent reliance on the idea that it is practically impossible for an arbitrageur to time their trades to coincide with the stock price reaching and leaving the reflecting boundary. Mathematically, an arbitrage is merely predictable process that satisfies certain technical conditions, and a model is arbitrage-free (in the appropriate sense) if no such process exists. The question of whether the strategy could be implemented in practice has no bearing on its existence as a well-defined mathematical object, and is thus irrelevant to the arbitrage properties of the model. Even though no trader could implement the strategy described in the proof of Proposition~\ref{propSec3:NIP}, it nevertheless exists as a well-defined predictable process satisfying the conditions in Definition~\ref{defSec2:NIP}.

\subsection{Failure of the Structure Condition}
Theorem~\ref{thmSec2:NIP<=>SC} showed that a continuous financial market model satisfies the NIP condition if and only if it satisfies the structure condition. Proposition~\ref{propSec3:NIP} proved that the RGBM model does not satisfy the NIP condition, by explicitly constructing an immediate profit strategy. Hence, the structure condition must fail as well. The next proposition proves this directly.

\begin{proposition}
\label{propSec3:SC}
The RGBM model does not satisfy the structure condition.	
\end{proposition}
\begin{proof}
Suppose that the structure condition does in fact hold. In that case, there is a predictable process $\vartheta$ satisfying \eqref{eqdefSec2:SC}. Using \eqref{eqSec3:PredChar}, that equation can be manipulated as follows:
\begin{align*}
&b_t(\omega)-\rho_t(\omega)=\vartheta_t(\omega)a_t^2(\omega)\\
\Rightarrow\qquad&b_t(\omega)\,\d G_t(\omega)-\rho_t(\omega)\,\d G_t(\omega)=\vartheta_t(\omega)a_t^2(\omega)\,\d G_t(\omega)\\
\Rightarrow\qquad&\mu\,\d t+\frac{\d L_t(\omega)}{b}-r\,\d t=\vartheta_t(\omega)\sigma^2\,\d t\\
\Rightarrow\qquad&\d\ell^b_t(\omega)=2\,\d L_t(\omega)=2b\bigl(\vartheta_t(\omega)\sigma^2-(\mu-r)\bigr)\,\d t,
\end{align*}
for $\P\otimes G\text{-a.a.}\;(\omega,t)\in\Omega\times\R_+$. This gives rise to a contradiction, since
\begin{equation*}
\begin{split}
0<\ell^b_{\tau_b(\omega)}(\omega)
&=\ell^b_{\tau_b(\omega)}(\omega)-\lim_{\varepsilon\downarrow 0}\ell^b_{\tau_b(\omega)-\varepsilon}(\omega)\\
&=\lim_{\varepsilon\downarrow 0}\int_{\tau_b(\omega)-\varepsilon}^{\tau_b(\omega)}\d\ell^b_t(\omega)\\
&=\lim_{\varepsilon\downarrow 0}\int_{\tau_b(\omega)-\varepsilon}^{\tau_b(\omega)}2b\bigl(\vartheta_t(\omega)\sigma^2-(\mu-r)\bigr)\,\d t
=0,
\end{split}
\end{equation*}
for $\P\text{-a.a}\;\omega\in\{\tau_b<\infty\}$, and $\P(\tau_b<\infty)>0$. Hence, the structure condition cannot hold.
\end{proof}

We can approach Proposition~\ref{propSec3:SC} from a slightly more intuitive angle. Using \eqref{eqSec3:A&M}, we obtain the following expressions for the continuous finite variation process $\hat{A}$ and the continuous local martingale $\hat{M}$, defined by \eqref{eqSec2:A^&B^}:
\begin{equation*}
\hat{A}_t=\int_0^t(\mu-r)\hat{S}_u\,\d u+\int_0^t\frac{\d L_u}{B_u}
\qquad\text{and}\qquad
\hat{M}_t=\int_0^t\sigma\hat{S}_u\,\d W_u,
\end{equation*}
for all $t\geq 0$, where $\hat{S}\coloneqq\sfrac{S}{B}$ is the discounted stock price. Since the local time of the stock price at the reflecting boundary increases instantaneously each time gets there, while the Lebesgue measure of the time spent by the stock price at the boundary is zero, we have
\begin{equation*}
\int_0^\infty\ind{\{S_t(\omega)=b\}}\,\d\ell^b_t(\omega)>0
\qquad\text{and}\qquad
\int_0^\infty\ind{\{S_t(\omega)=b\}}\,\d t=0,
\end{equation*}
for $\P\text{-a.a.}\;\omega\in\{\tau_b<\infty\}$.  Consequently,
\begin{align*}
\begin{split}
\int_0^\infty\ind{\{S_t(\omega)=b\}}\,\d\hat{A}_t(\omega)
&=\int_0^\infty\ind{\{S_t(\omega)=b\}}(\mu-r)b\,\d t+\int_0^\infty\ind{\{S_t(\omega)=b\}}\,\frac{\d L_t(\omega)}{B_t}\\
&=\int_0^\infty\ind{\{S_t(\omega)=b\}}\,\frac{\d\ell^b_t(\omega)}{2B_t}>0
\end{split}
\\
\intertext{and}
\int_0^\infty\ind{\{S_t(\omega)=b\}}\,\d\<\hat{M}\>_t(\omega)
&=\int_0^\infty\ind{\{S_t(\omega)=b\}}\sigma^2\frac{b^2}{B_t^2}\,\d t
=0,
\end{align*}
for $\P\text{-a.a.}\;\omega\in\{\tau_b<\infty\}$. Since $\P(\tau_b<\infty)>0$, it follows that $\hat{A}\centernot\ll\<\hat{M}\>$, in violation of the structure condition.\footnote{See the interpetation of the structure condition following Definition~\ref{defSec2:SC}.} In other words, the structure condition fails because a non-trivial set of  paths of $\hat{A}$ (those for which the first-passage time of the stock price to the reflecting boundary is finite) increase on sets of Lebesgue-measure zero (corresponding to the times when the stock price is at the reflecting boundary), while the paths of $\<\hat{M}\>$ cannot change on sets of Lebesgue-measure zero.

\subsection{Inconsistent Option Pricing Formulae}
Proposition~\ref{propSec3:NIP} established that the RGBM model does not satisfy the NIP condition. As a result, the NFLVR condition also fails, which rules out the existence of an equivalent risk-neutral probability measure. Operating under the mistaken belief that risk-neutral valuation is possible for the RGBM model, \citet{Vee08} derived pricing formulae for European options written on the stock. Subject to an amendment to the pricing formula for European put options by \citet{HV13}, these formulae have also been used by \citet{Her15} and \citet{HZ17}. Since their derivations are predicated on an incorrect assumption, we expect them to exhibit some inconsistencies.

Suppose we make the counterfactual assumption that an equivalent risk-neutral probability measure $\Prob{Q}\sim\P$ does in fact exist for the RGBM model. In that case, the discounted stock price process $\hat{S}$ is a $\Prob{Q}$-local martingale. Given a European contingent claim written on the stock, with maturity $T\geq 0$ and payoff function $\func{h}{[b,\infty)}{\R}$, we may (by hypothesis) apply a standard risk-neutral valuation approach to obtain its pricing function $\func{V^h}{[0,T]\times[b,\infty)}{\R}$, as follows:
\begin{equation*}
V^h(t,S)\coloneqq\E^{\Prob{Q}}_{t,S}\biggl(\frac{B_t}{B_T}h(S_T)\biggr)
=\E^{\Prob{Q}}_{t,S}\bigl(\e^{-r(T-t)}h(S_T)\bigr),
\end{equation*}
for all $(t,S)\in[0,T]\times[b,\infty)$. Here $\E^{\Prob{Q}}_{t,S}(\,\cdot\,)$ is the expected value operator with respect to the risk-neutral probability measure $\Prob{Q}_{t,S}$, under which $S_t=S\geq b$.

Consider European call and put options on the stock, with a common strike price of $K\geq b$ and a common maturity date $T\geq 0$. The pricing functions for these instruments are given by
\begin{align}
C(t,S)&\coloneqq\E^{\Prob{Q}}_{t,S}\bigl(\e^{-r(T-t)}(S_T-K)^+\bigr)\label{eqSec3:CallPrice-1}\\
\intertext{and}
P(t,S)&\coloneqq\E^{\Prob{Q}}_{t,S}\bigl(\e^{-r(T-t)}(K-S_T)^+\bigr),\label{eqSec3:PutPrice-1}
\end{align}
for all $(t,S)\in[0,T]\times[b,\infty)$. \citet{Vee08} derived a putative risk-neutral transition density for the stock price, which \citet{Vee08}, \citet{HV13} and \citet{Her15} used to evaluate \eqref{eqSec3:CallPrice-1} and \eqref{eqSec3:PutPrice-1}. In so-doing, they obtained the following pricing functions for European calls and puts:\footnote{The option pricing formulae are formulated differently (but equivalently) in the cited papers. We have chosen the formulations in \citet[Equations~(11) and (12)]{Her15}, which produce compact expressions.}
\begin{equation}
\label{eqSec3:CallPrice-2}
\begin{split}
C(t,S)&=S\Phi(z_1)-K\e^{-r(T-t)}\Phi\bigl(z_1-\sigma\sqrt{T-t}\bigr)\\
&\hspace{5em}+\frac{1}{\theta}\biggl(S\Bigl(\frac{b}{S}\Bigr)^{1+\theta}\Phi(z_3)-K\e^{-r(T-t)}\Bigl(\frac{K}{b}\Bigr)^{\theta-1}\Phi\bigl(z_3-\theta\sigma\sqrt{T-t}\bigr)\biggr)
\end{split}
\end{equation}
and
\begin{equation}
\label{eqSec3:PutPrice-2}
\begin{split}
P(t,S)&=K\e^{-r(T-t)}\Phi\bigl(-z_1+\sigma\sqrt{T-t}\bigr)-b\e^{-r(T-t)}\Phi(z_4)\\
&\hspace{5em}-S\bigl(\Phi\bigl(-z_4+\sigma\sqrt{T-t}\bigr)-\Phi(z_1)\bigr)\\
&\hspace{5em}-\frac{1}{\theta}\biggl(S\Bigl(\frac{b}{S}\Bigr)^{1+\theta}\Bigl(\Phi(z_4+\theta\sigma\sqrt{T-t}\bigr)-\Phi(z_3)\Bigr)-b\e^{-r(T-t)}\Phi(z_4)\\
&\hspace{10em}+K\e^{-r(T-t)}\Bigl(\frac{K}{b}\Bigr)^{\theta-1}\Phi\bigl(z_3-\theta\sigma\sqrt{T-t}\bigr)\biggr),
\end{split}
\end{equation}
for all $(t,S)\in[0,T]\times[b,\infty)$, where
\begin{align*}
&z_1\coloneqq\frac{\ln\frac{S}{K}+\bigl(r+\frac{1}{2}\sigma^2\bigr)(T-t)}{\sigma\sqrt{T-t}},
&z_3\coloneqq\frac{\ln\frac{b^2}{KS}+\bigl(r+\frac{1}{2}\sigma^2\bigr)(T-t)}{\sigma\sqrt{T-t}},\\
&z_4\coloneqq\frac{\ln\frac{b}{S}-\bigl(r-\frac{1}{2}\sigma^2\bigr)(T-t)}{\sigma\sqrt{T-t}},\qquad\text{and}
&\theta\coloneqq\frac{2r}{\sigma^2}.
\end{align*}
Here $\Phi(\,\cdot\,)$ denotes the cumulative distribution function for a standard normal random variable.

As an immediate consequence of \eqref{eqSec3:CallPrice-1}, we deduce that
\begin{equation}
\label{eqSec3:CallBnd}
C(t,S)\leq\E^{\Prob{Q}}_{t,S}\bigl(\e^{-r(T-t)}S_T\bigr)\leq S,
\end{equation}
for all $(t,S)\in[0,T]\times[b,\infty)$, where the second inequality follows because $\hat{S}$ is a supermartingale under $\Prob{Q}$, by virtue of Fatou's lemma.\footnote{\label{fnSec3:SMIneq}Since $\hat{S}$ is (by assumption) a local martingale under $\Prob{Q}$, which is bounded from below, an application of Fatou's lemma reveals that it is also a $\Prob{Q}$-supermartingale. Consequently,
\begin{equation*}
\E^{\Prob{Q}}\bigl(\e^{-r(T-t)}S_T\,|\,\SigAlg{F}_t\bigr)
=\e^{rt}\E^{\Prob{Q}}(\hat{S}_T\,|\,\SigAlg{F}_t)
\leq\e^{rt}\hat{S}_t
=S_t,
\end{equation*}
for all $t\geq 0$.} Note that \eqref{eqSec3:CallBnd} is a model-independent upper bound on call prices, whose violation contradicts the assumption that $\Prob{Q}$ is an equivalent risk-neutral probability measure. The next lemma reveals that this bound can be violated if call prices are determined by \eqref{eqSec3:CallPrice-2}.

\begin{proposition}
\label{propSec3:CallBnd}
The price of a call option obtained from \eqref{eqSec3:CallPrice-2} violates the upper bound \eqref{eqSec3:CallBnd} under certain parameter regimes, when the stock price is close to the reflecting boundary.
\end{proposition}
\begin{proof}
Suppose the model parameters satisfy $\theta=1$, which is to say that $r=\frac{1}{2}\sigma^2$. We consider the price of a call option at time $t\in[0,T)$ and assume that the stock price is located on the boundary $b$ at that time. In that case, \eqref{eqSec3:CallPrice-2} gives the following expression for the call price:
\begin{equation}
\label{eqpropSec3:CallBnd}
C(t,b)=2b\Phi\biggl(\frac{\ln\frac{b}{K}}{\sigma\sqrt{T-t}}+\sigma\sqrt{T-t}\biggr)-2K\e^{-r(T-t)}\Phi\biggl(\frac{\ln\frac{b}{K}}{\sigma\sqrt{T-t}}\biggr).
\end{equation}
Now, since
\begin{equation*}
\lim_{K\downarrow b}\Phi\biggl(\frac{\ln\frac{b}{K}}{\sigma\sqrt{T-t}}+\sigma\sqrt{T-t}\biggr)=\Phi\bigl(\sigma\sqrt{T-t}\bigr)>\frac{1}{2},
\end{equation*}
we can choose a small enough strike price $K'>b$ and a large enough risk-free interest rate $r'>0$ to ensure that
\begin{equation*}
\Phi\biggl(\frac{\ln\frac{b}{K'}}{\sigma\sqrt{T-t}}+\sigma\sqrt{T-t}\biggr)>\frac{1}{2}+\frac{K'}{b}\e^{-r'(T-t)}\Phi\biggl(\frac{\ln\frac{b}{K'}}{\sigma\sqrt{T-t}}\biggr).
\end{equation*}
Using those parameter values, \eqref{eqpropSec3:CallBnd} gives $C(t,b)>b$, in violation of \eqref{eqSec3:CallBnd}. The continuity of the call pricing function \eqref{eqSec3:CallPrice-2} with respect to the stock price ensures that the bound will also be violated for stock prices above the reflecting boundary, but sufficiently close to it.
\end{proof}

\begin{figure}
\includegraphics[scale=0.8]{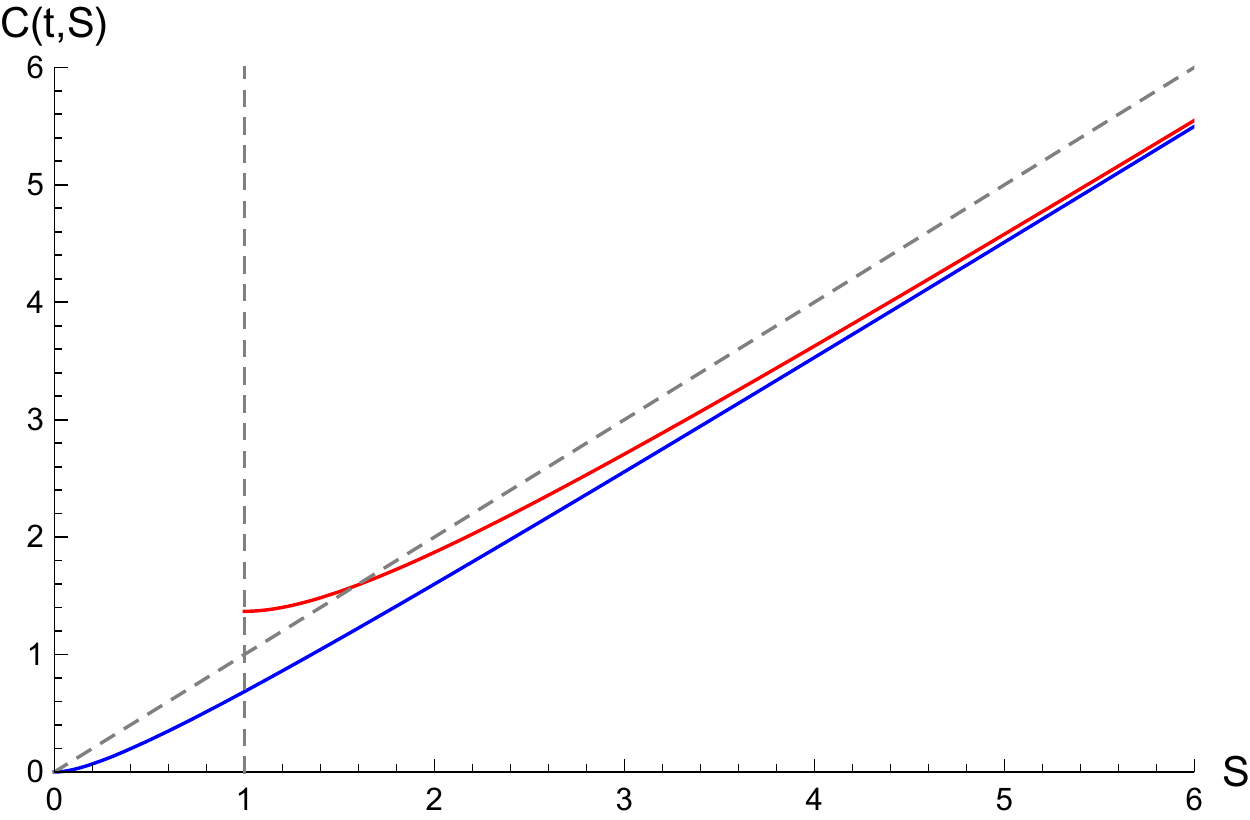}
\caption{The dependence of the RGBM call price \eqref{eqSec3:CallPrice-2} (solid red curve) and the \citet{BS73} call price (solid blue curve) on the stock price. The vertical dashed line is the reflecting boundary for the RGBM model and the sloped dashed line is the upper bound \eqref{eqSec3:CallBnd} for the call price. The parameter values are $T-t=10$, $K=2$, $r=0.125$, $\sigma=0.5$ and $b=1$.}
\label{figSec3:RGBM-Call}
\end{figure}

Proposition~\ref{propSec3:CallBnd} shows that call prices obtained from \eqref{eqSec3:CallPrice-2} can exceed the upper bound \eqref{eqSec3:CallBnd} when the stock price is close to the reflecting boundary. This is illustrated in Figure~\ref{figSec3:RGBM-Call}, which also illustrates that \citet{BS73} call prices are consistent with the upper bound. While the prices obtained from the two formulae converge as the stock price increases, we see that the RGBM call pricing formula \eqref{eqSec3:CallPrice-2} produces inflated prices that are inconsistent with risk-neutral valuation when the stock price approaches the reflecting boundary. This echoes our analysis of the failures of NIP and the structure condition, which revealed that the arbitrage pathologies of the RGBM model are intimately related to its reflecting behaviour.

We turn our attention now to put options. From \eqref{eqSec3:PutPrice-1}, we obtain
\begin{equation}
\label{eqSec3:PutBnd}
P(t,S)\geq\E^{\Prob{Q}}_{t,S}\bigl(\e^{-r(T-t)}(K-S_T)\bigr)\geq K\e^{-r(T-t)}-S,	
\end{equation}
for all $(t,S)\in[0,T]\times[b,\infty)$, where the second inequality follows because $\hat{S}$ is a supermartingale under $\Prob{Q}$, as previously noted. In this case, \eqref{eqSec3:PutBnd} imposes a model-independent lower bound on European put prices, whose violation once again contradicts the assumption that $\Prob{Q}$ is an equivalent risk-neutral probability measure. The next lemma reveals that this bound can be violated if put prices are determined by \eqref{eqSec3:PutPrice-2}.

\begin{proposition}
\label{propSec3:PutBnd}
The price of a put option obtained from \eqref{eqSec3:PutPrice-2} violates the lower bound \eqref{eqSec3:PutBnd} under certain parameter regimes, when the stock price is close to the reflecting boundary.
\end{proposition}
\begin{proof}
Suppose the model parameters satisfy $\theta=1$, which is to say that $r=\frac{1}{2}\sigma^2$. We consider the price of a put option at time $t\in[0,T)$ and assume that the stock price is located on the boundary $b$ at that time. In that case, \eqref{eqSec3:PutPrice-2} gives the following expression for the put price:
\begin{equation}
\label{eqpropSec3:PutBnd}
\begin{split}
P(t,b)&=K\e^{-r(T-t)}\biggl(1-2\Phi\biggl(\frac{\ln\frac{b}{K}}{\sigma\sqrt{T-t}}\biggr)\biggr)\\
&\hspace{5em}-2b\biggl(\Phi\bigl(\sigma\sqrt{T-t}\bigr)-\Phi\biggl(\frac{\ln\frac{b}{K}}{\sigma\sqrt{T-t}}+\sigma\sqrt{T-t}\biggr)\biggr).
\end{split}
\end{equation}
Now, since
\begin{equation*}
\Phi\bigl(\sigma\sqrt{T-t}\bigr)>\frac{1}{2}
\qquad\text{and}\qquad
\lim_{K\uparrow\infty}\Phi\biggl(\frac{\ln\frac{b}{K}}{\sigma\sqrt{T-t}}+\sigma\sqrt{T-t}\biggr)=0,
\end{equation*}
we can choose a large enough strike price $K'>b$ to ensure that
\begin{equation*}
\Phi\bigl(\sigma\sqrt{T-t}\bigr)-\Phi\biggl(\frac{\ln\frac{b}{K'}}{\sigma\sqrt{T-t}}+\sigma\sqrt{T-t}\biggr)>\frac{1}{2}.
\end{equation*}
Combining this with the fact that
\begin{equation*}
1-2\Phi\biggl(\frac{\ln\frac{b}{K'}}{\sigma\sqrt{T-t}}\biggr)<1
\end{equation*}
allows us to deduce from \eqref{eqpropSec3:PutBnd} that $P(t,b)<K'\e^{-r(T-t)}-b$, in violation of \eqref{eqSec3:PutBnd}. The continuity of the put pricing function \eqref{eqSec3:PutPrice-2} with respect to the stock price ensures that the bound will also be violated for stock prices above the reflecting boundary, but sufficiently close to it.
\end{proof}

\begin{figure}
\includegraphics[scale=0.8]{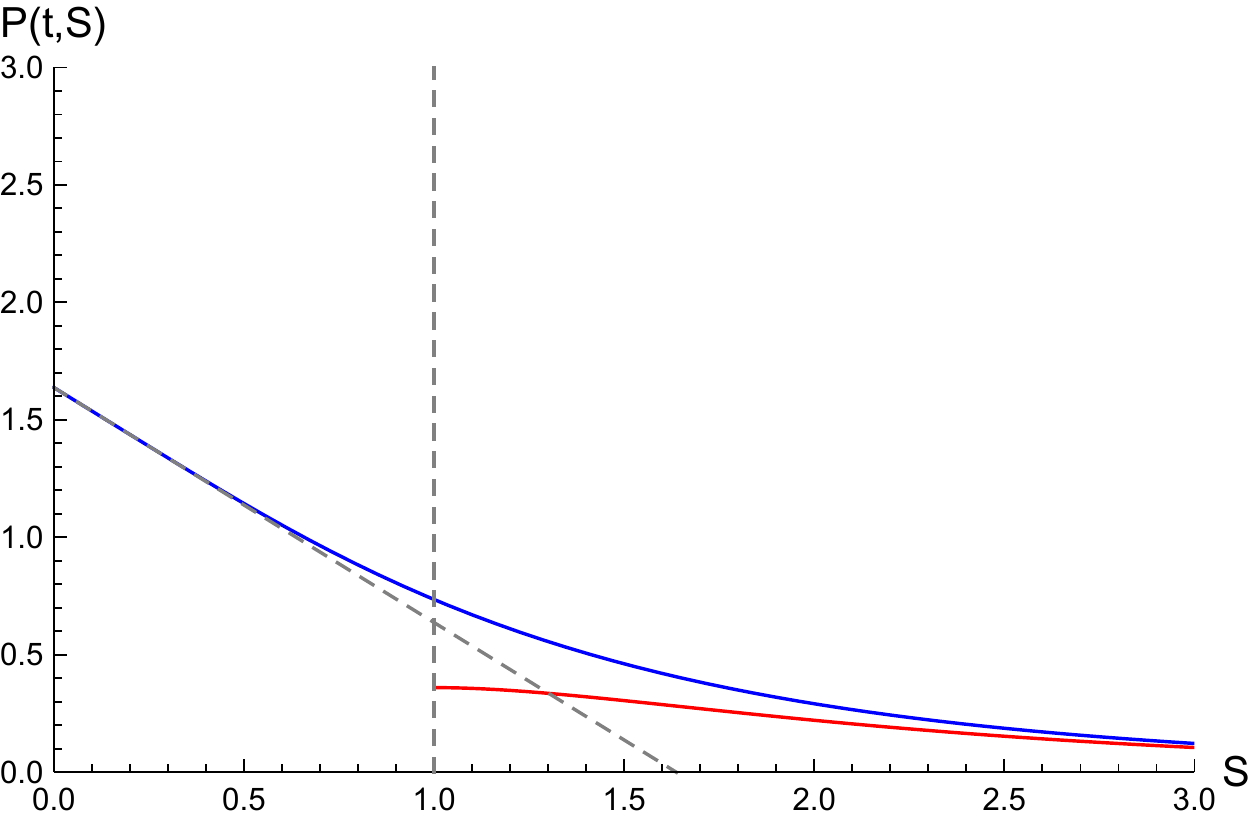}
\caption{The dependence of the RGBM put price \eqref{eqSec3:PutPrice-2} (solid red curve) and the \citet{BS73} put price (solid blue curve) on the stock price. The vertical dashed line is the reflecting boundary for the RGBM model and the sloped dashed line is the lower bound \eqref{eqSec3:PutBnd} for the put price. The parameter values are $T-t=10$, $K=2$, $r=0.02$, $\sigma=0.2$ and $b=1$.}
\label{figSec3:RGBM-Put}
\end{figure}

Proposition~\ref{propSec3:PutBnd} shows that put prices obtained from \eqref{eqSec3:CallPrice-2} can violate the lower bound \eqref{eqSec3:PutBnd} when the stock price is close to the reflecting boundary. This is illustrated in Figure~\ref{figSec3:RGBM-Put}, which also illustrates that \citet{BS73} put prices are consistent with the lower bound. While the prices obtained from the two formulae converge as the stock price increases, we see that the RGBM put pricing formula \eqref{eqSec3:PutPrice-2} produces diminished prices that are inconsistent with risk-neutral valuation when the stock price approaches the reflecting boundary. In other words, the pathologies of the RGBM model are once again evident near the reflecting boundary.

We conclude by emphasising that any violation of the bounds \eqref{eqSec3:CallBnd} and \eqref{eqSec3:PutBnd} is inconsistent with the assumption that an equivalent risk-neutral probability measure exists. Consequently, Propositions~\ref{propSec3:CallBnd}~and~\ref{propSec3:PutBnd} imply that either the option pricing formulae \eqref{eqSec3:CallPrice-2} and \eqref{eqSec3:PutPrice-2} are incorrect or the assumption that an equivalent risk-neutral probability measure exists is false.

\subsection{Problems with No-Negative-Equity Guarantees}
An \emph{equity release mortgage (ERM)} is a loan made to an elderly property-owning borrower that is collateralised by their property.\footnote{ERMS are typically known as \emph{reverse mortgages} outside the U.K.} In the U.K., ERMs typically embody a \emph{no-negative equity guarantee (NNEG)} stipulating that the amount due for repayment is no more than the minimum of the rolled-up loan amount and the property value at the time of repayment, which would be the time of the borrower’s death or their entry into a care home. This obligation to repay the minimum of two future values implies that a NNEG involves a put option issued to the borrower.

NNEG valuation plays a crucial role in the design and management of ERMs, and several valuation approaches have been considered.\footnote{See \citet{BD20} for a survey of NNEG valuation.} \citet{Tho21} recently proposed a new approach, built on the assumption that the government will intervene in the residential real estate market if property prices fall by more than a certain proportion, thereby establishing a de facto lower bound for house prices. Based on this idea, he proposed an RGBM model for house prices, where the lower reflecting boundary $b>0$ corresponds to the level at which the government will enter the property market to support prices. By rearranging the pricing formula for a European put on a dividend-paying security in \citet{Her15}, \citet{Tho21} obtained the following formula for the price of a NNEG with maturity date $T>0$ and loan principal $K\geq b$, written on a house whose current price is $S\geq b$:
\begin{equation}
\label{eqSec3:NNEGPrice}
\begin{split}
P(t,S)&=K\e^{-r(T-t)}\Phi\bigl(-z_1+\sigma\sqrt{T-t}\bigr)-S\e^{-q(T-t)}\Phi(-z_1)\\
&\hspace{5em}-b\e^{-r(T-t)}\Phi\bigl(-z_3+\sigma\sqrt{T-t}\bigr)+S\e^{-q(T-t)}\Phi(-z_3)\\
&\hspace{5em}+\frac{1}{\theta}\biggl(b\e^{-r(T-t)}\Phi\bigl(-z_3+\sigma\sqrt{T-t}\bigr)\\
&\hspace{10em}-S\e^{-q(T-t)}\Bigl(\frac{b}{S}\Bigr)^{1+\theta}\bigl(\Phi(z_4)-\Phi(z_2)\bigr)\\
&\hspace{10em}-K\e^{-r(T-t)}\Bigl(\frac{K}{b}\Bigr)^{\theta-1}\Phi\bigl(z_2-\theta\sigma\sqrt{T-t}\bigr)\biggr),
\end{split}
\end{equation}
for all $(t,S)\in[0,T]\times[b,\infty)$, where
\begin{align*}
z_1&\coloneqq\frac{\ln\frac{S}{K}+\bigl(r-q+\frac{1}{2}\sigma^2\bigr)(T-t)}{\sigma\sqrt{T-t}},
&z_2&\coloneqq\frac{\ln\frac{b^2}{KS}+\bigl(r-q+\frac{1}{2}\sigma^2\bigr)(T-t)}{\sigma\sqrt{T-t}},\\
z_3&\coloneqq\frac{\ln\frac{S}{b}+\bigl(r-q+\frac{1}{2}\sigma^2\bigr)(T-t)}{\sigma\sqrt{T-t}},
&z_4&\coloneqq\frac{\ln\frac{b}{S}+\bigl(r-q+\frac{1}{2}\sigma^2\bigr)(T-t)}{\sigma\sqrt{T-t}},\\
\intertext{and}
\theta&\coloneqq\frac{2(r-q)}{\sigma^2}.
\end{align*}
As before, $r\geq 0$ is the continuously compounding risk-free interest rate, while $\sigma>0$ is the volatility of the price of the house. The parameter $q\geq 0$, which is called the \emph{deferment rate}, is the continuously compounding discount rate that yields the \emph{deferment price} of the house, when applied to its current price.\footnote{The \emph{deferment price} of a house is the price payable now, for possession at some future date. Hence, the deferment rate $q$ may be regarded as a type of convenience yield.}

Since \eqref{eqSec3:NNEGPrice} is a straightforward extension of the European put pricing formula \eqref{eqSec3:PutPrice-2}, in order to accomodate a dividend-paying stock (with the deferment rate playing the role of a dividend yield), it shares the defects of that formula. First, the non-existence of an equivalent risk-neutral probability measure (or even a num\'eraire portfolio) for the RGBM model means that there is no economic justification for \eqref{eqSec3:NNEGPrice} as a pricing formula for NNEGs. Second, an easy modification of Proposition~\ref{propSec3:PutBnd} shows that \eqref{eqSec3:NNEGPrice} can violate the model-independent lower bound
\begin{equation}
\label{eqSec3:NNEGBnd}
P(t,S)\geq K\e^{-r(T-t)}-S\e^{-q(T-t)},
\end{equation}
for all $(t,S)\in[0,T]\times[b,\infty)$, when the house price is close to the reflecting boundary.\footnote{The lower bound \eqref{eqSec3:NNEGBnd} is a direct analogue of \eqref{eqSec3:PutBnd}, for the case of a dividend-paying stock. It can be derived in exactly the same way, if one assumes the existence of an equivalent risk-neutral probability measure.} As before, any violation of that bound contradicts the existence of an equivalent risk-neutral probability measure. To make matters worse, the next lemma highlights an additional defect of \eqref{eqSec3:NNEGPrice} that makes it especially ill-suited to NNEG valuation.

\begin{proposition}
\label{propSec3:NNEGBnd}
The price of a sufficiently long-dated NNEG obtained from \eqref{eqSec3:NNEGPrice} violates the lower bound \eqref{eqSec3:NNEGBnd}, if $r=0$ and $0<q<\frac{1}{2}\sigma^2$.
\end{proposition}
\begin{proof}
Suppose $r=0$ and $0<q<\frac{1}{2}\sigma^2$, in which case $-1<\theta<0$. Fix a current time $t\geq 0$ and a house price $S\geq b$, and observe that
\begin{align*}
-z_1+\sigma\sqrt{T-t}&=\BigO\Bigl(\frac{1}{\sqrt{T-t}}\Bigr)+\frac{1}{2}\sigma(1-\theta)\sqrt{T-t},
&-z_1&=\BigO\Bigl(\frac{1}{\sqrt{T-t}}\Bigr)-\frac{1}{2}\sigma(1+\theta)\sqrt{T-t},\\
-z_3+\sigma\sqrt{T-t}&=\BigO\Bigl(\frac{1}{\sqrt{T-t}}\Bigr)+\frac{1}{2}\sigma(1-\theta)\sqrt{T-t},
&-z_3&=\BigO\Bigl(\frac{1}{\sqrt{T-t}}\Bigr)-\frac{1}{2}\sigma(1+\theta)\sqrt{T-t},\\
z_4&=\BigO\Bigl(\frac{1}{\sqrt{T-t}}\Bigr)+\frac{1}{2}\sigma(1+\theta)\sqrt{T-t},
&z_2&=\BigO\Bigl(\frac{1}{\sqrt{T-t}}\Bigr)+\frac{1}{2}\sigma(1+\theta)\sqrt{T-t},\\
\intertext{and}
z_2-\theta\sigma\sqrt{T-t}&=\BigO\Bigl(\frac{1}{\sqrt{T-t}}\Bigr)+\frac{1}{2}\sigma(1-\theta)\sqrt{T-t},
\end{align*}
for all $T>t$. From this it follows that
\begin{align*}
\lim_{T\uparrow\infty}\Phi\bigl(-z_1+\sigma\sqrt{T-t}\bigr)&=1,&\lim_{T\uparrow\infty}\Phi(-z_1)&=0,\\
\lim_{T\uparrow\infty}\Phi\bigl(-z_3+\sigma\sqrt{T-t}\bigr)&=1,&\lim_{T\uparrow\infty}\Phi(-z_3)&=0,\\
\lim_{T\uparrow\infty}\Phi(z_4)&=1,&\lim_{T\uparrow\infty}\Phi(z_2)&=1,\\
\intertext{and}
\lim_{T\uparrow\infty}\Phi\bigl(z_2-\theta\sigma\sqrt{T-t}\bigr)&=1,
\end{align*}
by virtue of $-1<\theta<0$. Consequently, \eqref{eqSec3:NNEGPrice} gives
\begin{equation}
\label{eqpropSec3:NNEGBnd}
\lim_{T\uparrow\infty}P(t,S)=K-b+\frac{1}{\theta}\biggl(b-K\Bigl(\frac{K}{b}\Bigr)^{\theta-1}\biggr)
=K+\frac{b}{\theta}\biggl(1-\theta-\Bigl(\frac{K}{b}\Bigr)^\theta\biggr).
\end{equation}
The second term in this expression is negative, since $K>b>0$ and $\theta<0$, which implies that $\lim_{T\uparrow\infty}P(t,S)<K$. We may therefore choose $T'>t$ large enough, such that
\begin{equation*}
P(t,S)<K-S\e^{-q(T-t)},
\end{equation*}
for all $T>T'$, in violation of \eqref{eqSec3:NNEGBnd}.
\end{proof}

Proposition~\ref{propSec3:NNEGBnd} shows that NNEG prices obtained from \eqref{eqSec3:NNEGPrice} violate the lower bound \eqref{eqSec3:NNEGBnd} for sufficiently long-dated contracts, when the risk-free interest rate is zero and the deferment rate is not too large.\footnote{We could generalise the result by proving that long-dated NNEG prices obtained from \eqref{eqSec3:NNEGPrice} do not obey \eqref{eqSec3:NNEGBnd} for any parameter regimes with $-1<\theta<0$, but the proof would be messier.} Since NNEGs are typically long-dated instruments, the upshot is that they are dramatically undervalued by \eqref{eqSec3:NNEGPrice} in low interest rate environments.

\begin{figure}
\includegraphics[scale=0.8]{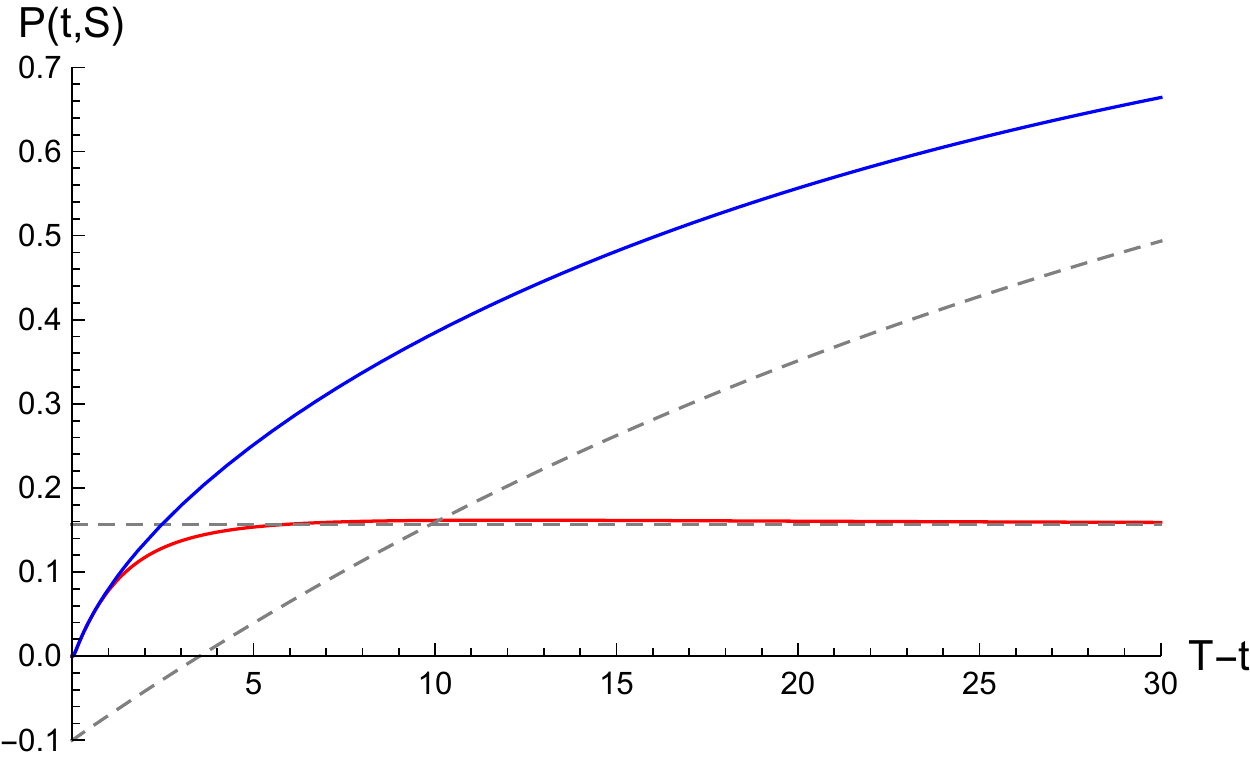}
\caption{The dependence of the RGBM NNEG price \eqref{eqSec3:NNEGPrice} (solid red curve) and the \citet{Bla76b} NNEG price (solid blue curve) on time to expiry. The horizontal dashed line is the long maturity asymptote \eqref{eqpropSec3:NNEGBnd} for the RGBM model and the dashed curve is the lower bound \eqref{eqSec3:NNEGBnd} for the price of the NNEG. The parameter values are $S=1$, $K=0.9$, $r=0$, $q=0.03$, $\sigma=0.3$ and $b=0.5$.}
\label{figSec3:RGBM-NNEG}
\end{figure}

Figure~\ref{figSec3:RGBM-NNEG} illustrates the issue described by Proposition~\ref{propSec3:NNEGBnd}. For the chosen parameter values, we see that \eqref{eqSec3:NNEGPrice} produces NNEG prices below the lower bound \eqref{eqSec3:NNEGBnd}, for maturities in excess of 10 years. We also see that the prices obtained from that formula converge asymptotically to the limit \eqref{eqpropSec3:NNEGBnd}. By contrast, \citet{Bla76b} NNEG prices are consistent with the lower bound \eqref{eqSec3:NNEGBnd}, for all maturities.\footnote{The Black (1976) put pricing formula is proposed by advocates of a market-consistent approach to NNEG valuation \citep[see][]{BD20}.} To appreciate the severity of the undervaluation issue for long-dated contracts, note that the \citet{Tho21} price of a 20-year NNEG in Figure~\ref{figSec3:RGBM-NNEG} is only $46\%$ of the value of the lower bound and a mere $29\%$ of the value of the corresponding \citet{Bla76b} NNEG price.

\section{Conclusions}
\label{Sec4}
Intuitively, an arbitrage is a trading strategy that realises a riskless profit without requiring an upfront investment of capital. But in the context of continuous-time finance, it is best to think of an arbitrage simply as a predictable process that satisfies certain technical conditions, with only a tenuous link to feasible real-world trading strategies. Moreover, there are several notions of arbitrage in continuous-time finance, with different implications for the mathematical features of a model. These observations warn us that intuition and heuristic reasoning are an unreliable substitute for mathematical rigour, when analysing the arbitrage properties of a continuous-time financial market model.

The salience of this warning is illustrated by the recent literature advocating the use of reflected geometric Brownian motion (RGBM) as a security price model. Several authors have argued that the RGBM model does not offer any arbitrage opportunities, but their argument is informal and heuristic. In particular, they do not specify what type of arbitrage opportunities are precluded and they do not formally verify the associated no-arbitrage condition. Instead, they appeal to an intuitive idea that arbitrageurs cannot profit from the reflecting behaviour of the security price in the model because the price process is continuous and the time spent on the reflecting boundary has Lebesgue measure zero. Based on these observations, they wrongly conclude that the RGBM model admits an equivalent risk-neutral probability measure, which leads them to derive invalid option pricing formulae.

A careful mathematical analysis of the RGBM model shows that it violates even the weakest no-arbitrage condition considered in the literature. Consequently, it does not offer any of the structural features required for contingent claim valuation. In particular, it does not admit a num\'eraire portfolio or an equivalent risk-neutral probability measure. As a result, there is no theoretical justification for the published RGBM option pricing formulae. Indeed, a close examination of those formulae reveals that they behave pathologically under certain conditions.


\bibliography{ProbFinBiblio}
\bibliographystyle{chicago}
\end{document}